%% file: SimpCharacsOfTSNets.tex
\documentclass[aps,pre,superscriptaddress,groupedaddress,preprint]{revtex4}  
\usepackage{epsfig}
\usepackage{amsmath}
\usepackage{multirow,bigdelim}
\usepackage{booktabs} 

\usepackage{graphicx}        
\usepackage{subfig}
\usepackage{caption}
\DeclareCaptionFormat{center}

\usepackage{array}
\newcolumntype{C}[1]{>{\centering\let\newline\\\arraybackslash\hspace{0pt}}m{#1}}
\begin{document}
\title{The simplicial characterisation of TS networks: Theory and applications}

\input author_list.tex


\begin{abstract}
We use the visibility algorithm to construct the time series networks obtained from the time series of different dynamical regimes of the logistic map. We define the simplicial characterisers of networks which can analyse the simplicial structure at both the global and local levels. These characterisers are used to analyse the TS networks obtained in different dynamical regimes of the logisitic map. It is seen that the simplicial characterisers are able to distinguish between distinct dynamical regimes. We also apply the simplicial characterisers to time series networks constructed from fMRI data, where the preliminary results indicate that the characterisers are able to differentiate between distinct TS networks.
\end{abstract}


\maketitle

\section{Introduction}
\label{sec:1}

The analysis of time series of evolving dynamical systems is a well established area of research. There are numerous well developed techniques for the analysis 
of such time series. These include Fourier transforms, power spectra, dimensions and entropies, Lyapunov exponents etc. These characterisers provide valuable insights into the dynamical behaviours of the evolving systems. In recent years, new techniques have emerged for the analysis of time series. These consist of mapping the time series to networks, using a variety of algorithms such as the visibility algorithms, recurrence times, identification of cycles or correlations, etc. See  \cite{TSnetreview} for a brief review.  
Since networks are also a well established paradigm in the study of complex systems, there are well established metrics for their analysis. These include path lengths, clustering co-efficients, degree distributions etc. Here, we introduce a series of network characterisers which go beyond these usual characterisers, and provide new insights into the dynamical behaviour of systems. The characterisers are based on the methods of algebraic topology. We demonstrate the utility of these characterisers in the time series arising from the logistic map, and demonstrate that the characterisers can differentiate between different dynamical regimes. We also analyse the time series obtained from the fMRI analysis of neural data to demonstrate the general applicability of the method.

\section{The Visibility Graph}

The visibility algorithm for converting a time series into an equivalent network, was proposed by Lacasa and Lucque \cite{LL}. We note that there are other methods for converting the time series graphs into networks, such as recurrence algorithms etc. These are summarised in \cite{TSnetreview}. We use the visibility algorithm here on account of its intuitive nature. 
The visibility algorithm is implemented by connecting two points $ (y_i, t_i)$
and $(y_j,t_j)$ by a straight line, provided no other intermediate point, $(y_r
, t_r)$ lies above the line, i.e. $(y_i,t_i)$ and $(y_j,t_j)$ should be `visible' to each other, with no other intermediate point obstructing the line of visibility in between (See Fig. 1). For this, all intermediate points should satisfy the condition

\begin{equation}
	y_j > y_r + \frac{y_j - y_i}{t_j - t_i} (t_j - t_r). \\
\end{equation}

It has been demonstrated that the visibility algorithm is capable of capturing series correlations (such as periodicity, fractality and chaoticity) and has been used in diverse contexts from geophysics \cite{geo} to finance \cite{finance}.
The graphs so generated have been analysed using the usual characterisers, such as degree distributions, clustering co-efficients, average path lengths and Hurst exponents. 
Our aim is to analyse these graphs using new simplicial characterisers, which go beyond the usual network analysis.

The specific time series that we use is the time series generated from the logistic map $x_{n+1} = \mu x_n (1 - x_n)$, with the parameter $\mu$ lying in the interval $[0,4]$ and $x_n \in [0,1]$. These time series are shown for the values 
$\mu=3.566 $ (period 16) and $\mu= 3.56995$ (edge of chaos)
in Figs 1(a) and 1(b). The corresponding time series networks are also shown in Figs 2(a) and 2(b). We note that the network representation at the periodic value $\mu= 3.566 $ shows the underlying periodicity in the repetition of the connection pattern, whereas the $\mu=3.56995$ (edge of chaos) shows a much more irregular behaviour. Similar network representations have been obtained in \cite{LL}, but have not been analysed further quantitatively.
Here, we analyse the TS graphs obtained in Figs 2(a) and 2(b) using the methods of algebraic topology. The relevant simplicial characterisers are defined in the next section.

\begin{figure*}[!h]
	\begin{center}
		\begin{tabular}{c}
			\includegraphics[width=0.65\textwidth]{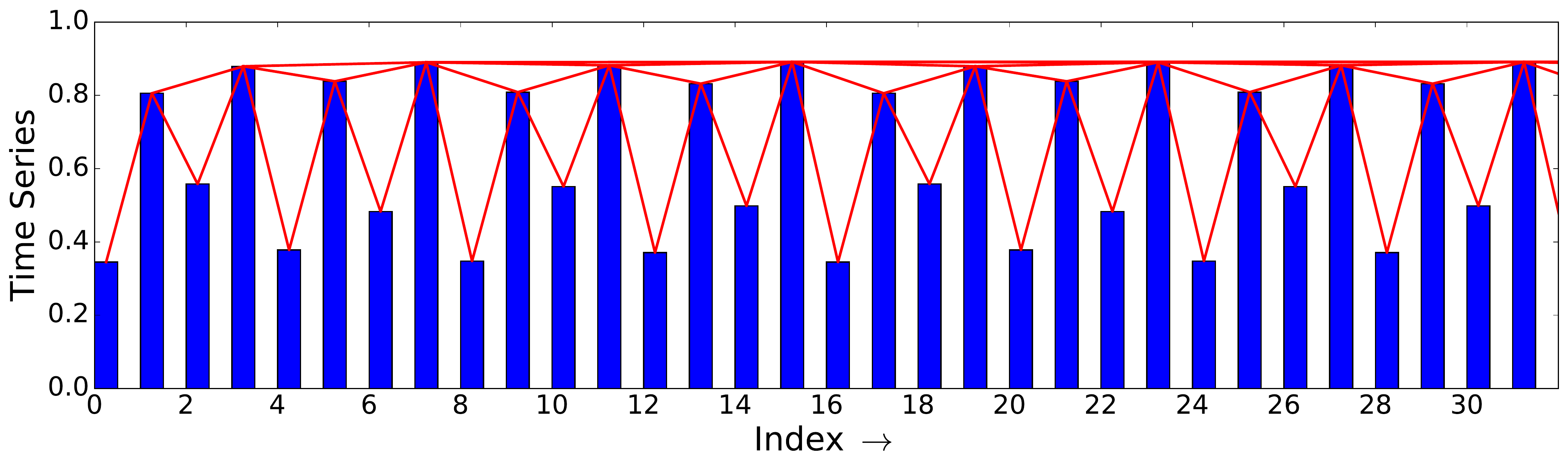}\\
			(a)\\
			\includegraphics[width=0.65\textwidth]{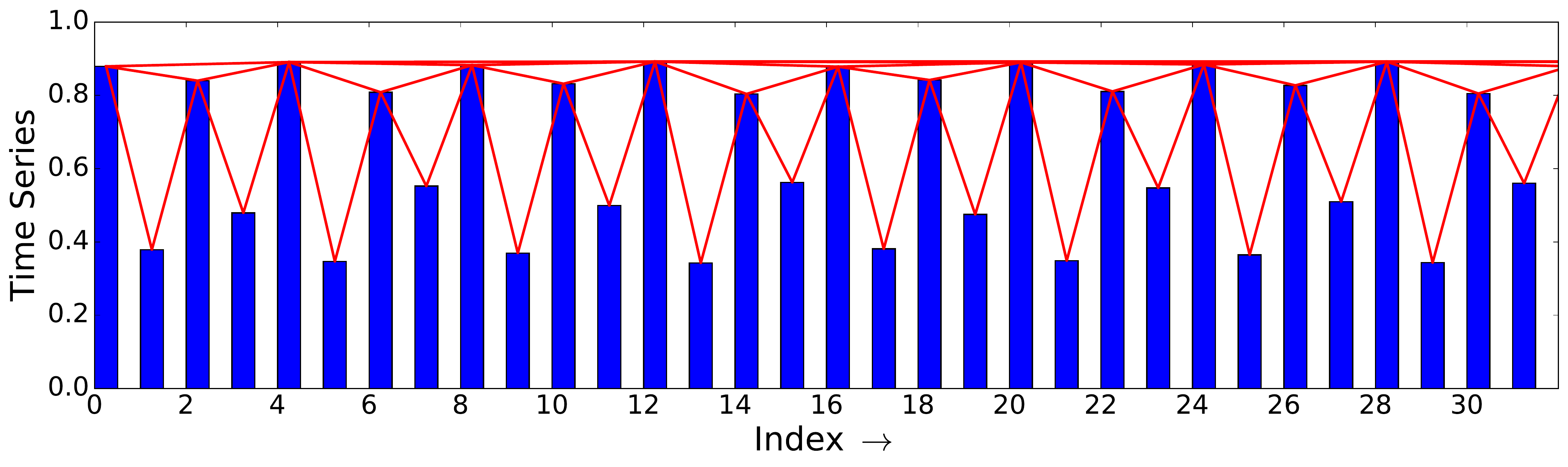}\\
			(b)\\
		\end{tabular}{}
		\caption{Portion of logistic map time series with visibility connections for (\textbf{a}) period 16 ($\mu$ = 3.566), and (\textbf{b}) edge of chaos ($\mu$ = 3.56995). Number of points shown is 32.}
		\label{fig:TS}
	\end{center}
\end{figure*}

\begin{figure*}[!h]
	\begin{center}
		\begin{tabular}{cc}
			\includegraphics[width=0.35\textwidth]{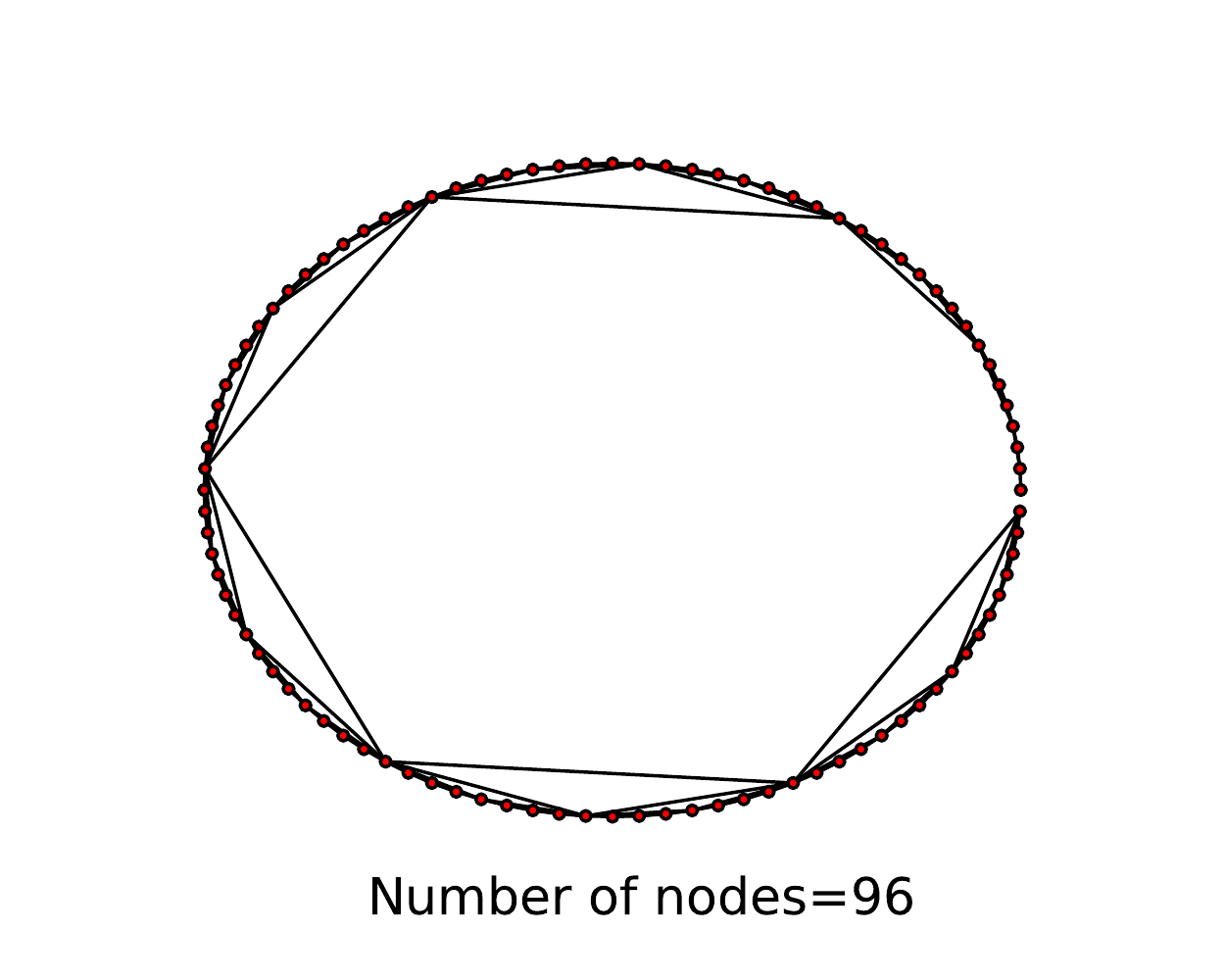}
			&
			\includegraphics[width=0.35\textwidth]{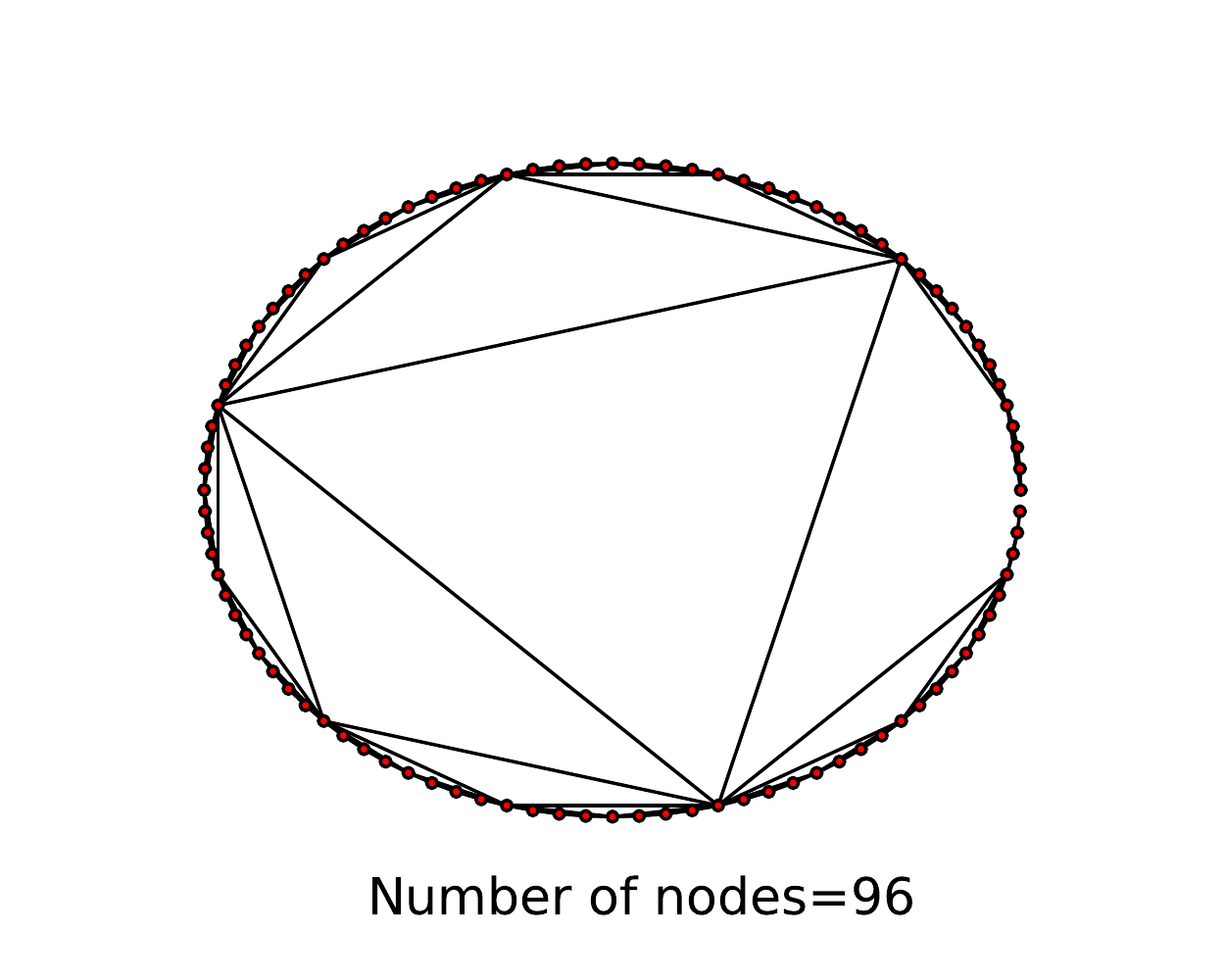}\\
			(a) & (b)\\
		\end{tabular}{}
		\caption{Corresponding TS-networks for (\textbf{a}) period 16 ($\mu$ = 3.566), and (\textbf{b}) edge of chaos ($\mu$ = 3.56995). Number of nodes is 96.}
		\label{fig:networks}
	\end{center}
\end{figure*}

\label{sec:2}
\section{The definitions of the simplicial characterisers}

The simplicial characterisers defined here can be used to analyse any graph or network. Here, a graph or network, is defined to be a collection of nodes interacting via interconnected edges or links. We define a clique to be a maximal complete subgraph, i.e. the nodes of a clique are not part of a larger complete sub-graph. Using the adjacency matrix of a the network, the Bron-Kerbosch algorithm \cite{Bron} is used to identify the cliques. The cliques are regarded as the simplicial complexes of the graph. 

A simplex with $q+1$ nodes or vertices, is a $q$ dimensional simplex. If two simplices have $q+1$ nodes in common, they share a $q$ face. A collection of simplices, i.e. the nodes and the shared faces form a simplicial complex. We are interested in the $q-$connectedness of the simplex, as well as in the dimension of the simplicial complex, i.e. the dimension of the largest simplex in the complex. If we can find a sequence of simplices such that each successive pair share a $q$ face, then all the simplices in this sequence are said to be $q-$connected. Simplices which are $q-$connected, are also connected at all lower levels.  

We define six simplicial characterisers, both global and local. Three of these quantities are well known, and defined in most algebraic topology texts \cite{jonsson}, and three  are new and have been recently defined in the context of social and traffic networks \cite{Maletic, Andjelkovich}. 

\begin{enumerate}

\item The first structure vector $\mathbf{Q} = Q_{0},Q_{1}, \ldots Q_{qmax}$: 
The $q-$th component of the vector is the number of $q-$ connected components at the $q$ th level.
This is a measure of the connectivity of the clique complex at various levels.   
\item The next quantity is an auxiliary vector, which is denoted by $\mathbf{\tilde{f}}$ and has the number of $q-$ dimensional simplices as its $q-$th component. This vector has been observed to behave as a response function, in the analysis of traffic \cite{Andjelkovich}.

\item The second structure vector: $\mathbf{N_s}=n_0, n_1, \ldots, n_{qmax}$. This vector has the the number of simplices of dimension $q$ and higher, as its $q-$ th component. This vector is thus related to the auxilary vector $\mathbf{\tilde{f}}$, in the sense that its components are a running sum of the components of $\mathbf{\tilde{f}}$.

\item The third structure vector $\mathbf{\widehat{Q}}$: This is constructed out of the components of the first two structure vectors. Its $q-$th component is defined as 
$\widehat{Q}_q= \left(1-\frac{Q_q}{n_q}\right)$.

\item   

The topological dimension of a node $i$  $\mathrm{dim}\, Q^{i}$: 

The topological dimension of node $i$ of the simplicial complex, is given by

\begin{equation}
\mathrm{dim}\,Q^i = \sum_{q=0}^{q_{\mathrm{max}}}\,Q_q^i,
\end{equation}

where $q_{\mathrm{max}}$ is the dimension of the simplicial complex, and $Q_{k}^{i}$ is the number of different simplices of dimension $k$ in which the node $i$ participates.

\item The topological entropy $\mathbf{S}$: This is defined by the equation 

\begin{equation}
S_Q(q) = - \frac{\sum_i\, p_q^i\,\mathrm{log}\,p_q^i}{\mathrm{log}\,N_q}.
\end{equation}

were, $p_q^i = {Q_q^i}/{\sum_i\, Q_q^i} $ is defined to be the probability that a given  node $i$ participates  in a $q$-simplex, and $N_q = \Sigma_i \left(1 - \delta_{Q_q^i,0}\right)$ is the number of nodes that participate in at least one $q$-simplex.

\end{enumerate}

We note that there are six quantities, five of which are global, except for the fifth quantity, viz. the topological dimension of a node $i$  $\mathrm{dim}\, Q^{i}$, which is a local quantity, which turns out to be of maximum utility in identifying different dynamical regimes.  
The three structure vectors $\mathbf{Q}$, $\mathbf{N_s}$, $\mathbf{\widehat{Q}}$  
are well known and have been defined earlier for simplicial analysis \cite{jonsson}. As mentioned above, the remaining quantities were first defined in the context of social and traffic networks \cite{Maletic,Andjelkovich}.
We now apply these quantities to the analysis of the logistic map time series.

%
\section{The simplicial analysis of the logistic map time series}

\label{subsec:2}

We obtain the time series of the logistic map at the parameter values $\mu=3.566 $ (period-16) and $\mu= 3.56995$ (edge of chaos) and construct the equivalent networks
and characterise them using the simplicial characterisers. The resulting values for the six topological characterisers are listed in Tables 1, 2, and 3.

The first interesting fact is the number of levels in the network.
We note that the network contains simplices at the $q=0$, $q=1$ and $q=2$ levels.
We note that there are no isolated points, for either parameter value.
At the $q=1$ level, the period-$16$ network contains $628$ $1-$connected components, whereas the network at the edge of chaos contains $19$ components at this level.
At the $q=2$ level, the period-$16$ has $9371$ $2-$connected components, whereas the edge of chaos network has $9980$ connected components at this level.
Thus the edge of chaos network has more components which are connected at the higher level.
The $\mathbf{\tilde{f}}$ vector which counts the number of $q-$ dimensional simplices at the $q-$th level, also shows more simplices at the $q=2$ level for the edge of chaos network (9980) than the period-16 network (9371).
Thus, the network at the edge of chaos is more connected than the period-16 network.
This behaviour is also reflected in the entropies where the edge of chaos network has a lower entropy (0.96210) than the period-16 network (0.96864).
However, the quantity which picks up the difference between the two networks most sharply is the $\mathrm{dim}\, Q^{i}$ which is the maximum value of the topological dimension of all the nodes in the network.
For the period-16 network, this is $\mathrm{dim}\, Q^{i} = 8$ whereas the edge of chaos network has $\mathrm{dim}\, Q^{i}=23$, a significantly higher value.
Thus the most connected node participates in a much higher number of simplices.
We therefore conclude that the higher the chaoticity of the dynamical states, the more interconnected are their networks.
There is thus a direct and quantifiable connection between the correlations in the dynamical state, and the simplicial structure at all levels.
We note that simplicial analysis is capable of detecting the nature of the dynamical state in other systems as well.

\begin{table}[h!]
\centering
\caption{Structure vectors $\mathbf{Q}$, $\mathbf{N_{s}}$ and $\mathbf{\widehat{Q}}$ for the TS networks of the logistic map at parameter values of $\mu$ = 3.566 (period 16) and $\mu$ = 3.56995 (edge of chaos). The time series considered is of length 10000.}
\begin{tabular}{| p{1.5cm} | C{1.5cm} | C{1.5cm} | C{1.5cm} | C{1.5cm} | C{1.5cm} | C{1.5cm} |}
\hline
	&	\multicolumn{3}{l |}{$\mu$ = 3.566 (period 16).}					&	\multicolumn{3}{l |}{$\mu$ = 3.56995 (edge of chaos).}						\\	\hline
$q$-level	&	$\mathbf{Q}$	&	$\mathbf{N_{s}}$	&	$\mathbf{\widehat{Q}}$	&	$\mathbf{Q}$	&	$\mathbf{N_{s}}$	&	$\mathbf{\widehat{Q}}$		\\	\hline
0	&	1	&	9372	&	0.99989	&	1	&	9981	&	0.99990	\\	\hline
1	&	628	&	9372	&	0.93299	&	19	&	9981	&	0.99810	\\	\hline
2	&	9371	&	9371	&	0	&	9980	&	9980	&	0	\\	\hline
\end{tabular}
\label{table:topol_charac_values_1}
\end{table}
\begin{table}[h!]
\centering
\caption{Structure vector $\mathbf{\tilde{f}}$ and entropy $\mathbf{S}$ for the TS networks of the logistic map at parameter values of $\mu$ = 3.566 (period 16) and $\mu$ = 3.56995 (edge of chaos). The time series considered is of length 10000.}
\begin{tabular}{| p{1.5cm} | C{1.5cm} | C{1.5cm} | C{1.5cm} | C{1.5cm} |}
\hline
	&	\multicolumn{2}{l |}{$\mu$ = 3.566 (period 16).}				&	\multicolumn{2}{l |}{$\mu$ = 3.56995 (edge of chaos).}		\\	\hline
q-level	&	$\mathbf{\tilde{f}}$	&	$\mathbf{S}$	&	$\mathbf{\tilde{f}}$	&	$\mathbf{S}$	\\	\hline
0	&	0	&	0	&	0	&	0	\\	\hline
1	&	1	&	1	&	1	&	1	\\	\hline
2	&	9371	&	0.96864	&	9980	&	0.96210	\\	\hline
\end{tabular}
\label{table:topol_charac_values_2}
\end{table}
\begin{table}[h!]
\centering
\caption{Maximum value of the topological dimension of all nodes in the TS network of the logistic map at parameter values of $\mu$ = 3.566 (period 16) and $\mu$ = 3.56995 (edge of chaos). The time series considered is of length 10000.}
\begin{tabular}{| p{4cm} | C{2.5cm} | }
\hline
\multicolumn{1}{| c |}{$\mu$}	&	max(dim $Q^{i}$)	\\	\hline
3.566 (period 16)	&	8	\\	\hline
3.56995 (edge of chaos)	&	23	\\	\hline
\end{tabular}
\label{table:topol_charac_values_3}
\end{table}

\begin{table*}[!t]
\centering
\caption{Values of simplicial characterisers max(dim Q$^{i}$ ) and $\mathbf{S}$ for the fMRI data.
EW = English word tasks, 
ER = English rest tasks, 
ENW = English non-word tasks, 
HW = Hindi word tasks, 
HR = Hindi rest tasks, and
HNW = Hindi non-word tasks.}
\begin{tabular}{c}
	Table 4a: ROI Left Angular Gyrus for the Adult.\\
	\begin{tabular}{ | p{1.5cm} | C{2.5cm} | C{2.5cm} | C{2.5cm} | C{2.5cm} | }
		\hline
		Sub-task	&	max(dim $Q^{i}$)	&	S(0)	&	S(1)	&	S(2)	\\	\hline
		EW	&	7	&	0	&	0.9784957361	&	0.9515456317	\\	\hline
		ER	&	9	&	0	&	1	&	0.9654711418	\\	\hline
		ENW	&	7	&	0	&	0.9881090916	&	0.9622363428	\\	\hline
		HW	&	10	&	0	&	0	&	0.9608581361	\\	\hline
		HR	&	8	&	0	&	0.9788379142	&	0.9639586918	\\	\hline
		HNW	&	8	&	0	&	0.9788379142	&	0.9615663021	\\	\hline
	\end{tabular}
	\\
	Table 4b: ROI Left Calcarine Sulcus for the Adult.\\
	\begin{tabular}{ | p{1.5cm} | C{2.5cm} | C{2.5cm} | C{2.5cm} | C{2.5cm} | }
		\hline
		Sub-task	&	max(dim $Q^{i}$)	&	S(0)	&	S(1)	&	S(2)	\\	\hline
		EW	&	8	&	0	&	0.9775626836	&	0.9548140045	\\	\hline
		ER	&	11	&	0	&	0.9848586618	&	0.9602719488	\\	\hline
		ENW	&	6	&	0	&	0.9784957361	&	0.9677737123	\\	\hline
		HW	&	8	&	0	&	0.9697238999	&	0.9595191482	\\	\hline
		HR	&	9	&	0	&	0.9697238999	&	0.9651100304	\\	\hline
		HNW	&	6	&	0	&	0.9739760316	&	0.9618112835	\\	\hline
	\end{tabular}
\end{tabular}
\end{table*}
\begin{table*}[!t]
\centering
\begin{tabular}{c}
	Table 4c: ROI Left Angular Gyrus for the Child.\\
	\begin{tabular}{ | p{1.5cm} | C{2.5cm} | C{2.5cm} | C{2.5cm} | C{2.5cm} | }
		\hline
		Sub-task	&	max(dim $Q^{i}$)	&	S(0)	&	S(1)	&	S(2)	\\	\hline
		EW	&	6	&	0	&	0.9695703502	&	0.9619605169	\\	\hline
		ER	&	12	&	0	&	0.9671320181	&	0.96011217	\\	\hline
		ENW	&	8	&	0	&	0.9823368126	&	0.9479697481	\\	\hline
		HW	&	8	&	0	&	0.9784957361	&	0.9532572532	\\	\hline
		HR	&	8	&	0	&	0.9823368126	&	0.9593979771	\\	\hline
		HNW	&	8	&	0	&	0.9767405285	&	0.9550307474	\\	\hline
	\end{tabular}
	\\
	Table 4d: ROI Left Calcarine Sulcus for the Child.\\
	\begin{tabular}{ | p{1.5cm} | C{2.5cm} | C{2.5cm} | C{2.5cm} | C{2.5cm} | }
		\hline
		Sub-task	&	max(dim $Q^{i}$)	&	S(0)	&	S(1)	&	S(2)	\\	\hline
		EW	&	8	&	0	&	0.9796133098	&	0.9647246463	\\	\hline
		ER	&	9	&	0	&	0.9795697645	&	0.9608848295	\\	\hline
		ENW	&	8	&	0	&	0.9795697645	&	0.9642018687	\\	\hline
		HW	&	10	&	0	&	0.9823368126	&	0.9561739848	\\	\hline
		HR	&	8	&	0	&	0.9766874637	&	0.9643463718	\\	\hline
		HNW	&	7	&	0	&	0.9796133098	&	0.9585418732	\\	\hline
	\end{tabular}
\end{tabular}
\end{table*}

%
\section{Simplicial analysis of fMRI data}
\label{sec:3}

In order to demonstrate the potential  of  simplicial analysis, we demonstrate its application to fMRI (Functional Magnetic Resonance Imaging) data. The data is taken from two regions of imaging (ROI), the left angular gyrus, and the left calcarine sulcus. The first region is associated with complex language functions (i.e. reading, writing and interpretation of what is written), and the second region is where the primary visual cortex is concentrated. There are two subjects, one adult and one child, each of whom is carrying out a reading task. The importance of the two regions of imaging for the reading task is obvious.

Each subject is carrying out a reading task from a screen in two distinct languages, English and Hindi, with Hindi being the subjects' native language. The time series recorded has 480 points recorded  as follows: the subject reads words in the given language (20 data points), followed by a rest period (20 data points), then non-words in the language (e.g. `cart' and `rarn', for English) (20 data points) followed by another rest. Three repeats of each sequence are carried out, for each language, English, followed by Hindi.

The values of the simplicial characterisers max(dim Q$^{i}$) and $\mathbf{S}$ are shown in Tables 4a, 4b, 4c, and 4d. It is clear that the data shows detailed variations between the two ROIs, the two languages for each subject, and also between subjects of different ages. While the analysis is too preliminary for definite conclusions, it is clear that this method of analysis is capable of yielding insights which are not accessible by other methods.

%
\section{Conclusions}
\label{sec:3}

To summarise, we propose the simplicial characterisers of TS networks. These characterisers are capable of analysing the TS networks at each level of simplicial structure and hence can provide a detailed analysis of the correlations in the underlying time series data. We demonstrate the utility of these characterisers in the context of the well known logistic map, where the dynamical regimes are very well understood. The simplicial characterisers turn out to be capable of distinguishing between the dynamical regimes, and also provide insights into the detailed structure of the TS networks. We also demonstrate the application of these characterisers to fMRI data. Here, the analysis is preliminary and no definite conclusions can be drawn. However, the fact that the characterisers are capable of distinguishing between different regimes, and subjects of different ages, demonstrates the potential of the method. We hope these characterisers will turn out to be useful in diverse application contexts.

%
\section{Acknowledgements}
We would like to thank Dr.\,Nandini Chatterjee Singh and Dr.\,Sarika Cherodath of NBRC, Manesar for the fMRI data, and N.\,Nithyanand Rao for earlier collaboration.

%
\bibliographystyle{apsrev}
\bibliography{refs}        

\end{document}

%% file: author_list.tex
\author{Neelima Gupte}
\email{gupte@physics.iitm.ac.in}
\affiliation{Department of Physics, Indian Institute of Technology Madras, Chennai - 600 036, India}

\author{N. Nirmal Thyagu}
\email{nirmalthyagu@gmail.com}
\affiliation{Vellore Institute of Technology, Chennai - 600 127, India}

\author{Malayaja Chutani}
\email{malayajac@physics.iitm.ac.in}
\affiliation{Department of Physics, Indian Institute of Technology Madras, Chennai - 600 036, India}